\documentclass[prl, aps, amsfonts, twoside, twocolumn, amssymb, superscriptaddress, sort&compress, reprint]{revtex4-1}

\usepackage[latin1]{inputenc}
\usepackage[UKenglish]{babel}
\usepackage{lmodern, graphicx, amsmath, textcomp, xcolor, mathtools, siunitx, cancel, nicefrac}

\setcitestyle{super}

\newcommand{\ket}[1]{\lvert#1\rangle}
\newcommand{\ops}[1]{\hat{#1}}
\newcommand{\photonum}{\ensuremath{\expct{\ops n}}}

\newcommand{\eqn}{Eq.}
\newcommand{\fig}{Fig.}

\DeclarePairedDelimiter\expct{\langle}{\rangle}
\DeclarePairedDelimiter\robra{\lparen}{\rparen}

\begin{document}

\title{Mutual information of optical communication in phase conjugating Gaussian channels}

\author{Clemens Sch\"afermeier}
\email{clemens@fh-muenster.de}
\affiliation{Center for Macroscopic Quantum States (bigQ), Department of Physics, Technical University of Denmark, 2800 Kgs.\ Lyngby, Denmark}
\affiliation{Kavli Institute of Nanoscience, Delft University of Technology, 2628 CJ Delft, The Netherlands}

\author{Ulrik L.\ Andersen}
\email{ulrik.andersen@fysik.dtu.dk}
\affiliation{Center for Macroscopic Quantum States (bigQ), Department of Physics, Technical University of Denmark, 2800 Kgs.\ Lyngby, Denmark}

\begin{abstract}
In all practical communication channels, the code word consist of Gaussian states and the measurement strategy is often a Gaussian detector such as homodyning or heterodyning.
In this paper we investigate the communication performance using a phase-conjugated alphabet and joint Gaussian detection in a phase-insensitive amplifying channel.
We find that a communication scheme consisting of a phase-conjugated alphabet of coherent states and a joint detection strategy significantly outperforms a standard coherent-state strategy based on individual detection.
Moreover, we show that the performance can be further enhanced by using entanglement and that the performance is completely independent on the gain of the phase-insensitively amplifying channel.
\end{abstract}

\maketitle

\section{Introduction}

The classical channel capacity of a bosonic quantum channel plays a ubiquitous role both in optical classical communication as well as in quantum key distribution.
The capacity determines the maximal communication rate and is strongly related to the maximally achievable secure key rate \cite{Cover2006, Scarani2009}.
Very encouragingly, it has been found that the capacity -- equivalent to the maximal mutual information -- can be reached in a lossy and noisy bosonic channel by using readily available coherent states of light \cite{Giovannetti200401, Giovannetti2014}.
On the downside, however, it also requires a receiver that jointly detects long code words by means of highly non-linear transformations which are currently not practical.
In most optical communication realisations, the receiver is Gaussian, which means that it transforms a Gaussian alphabet of coherent states into Gaussian detector statistics.
Prominent examples of Gaussian receivers are homodyne and heterodyne detectors potentially combined with Gaussian quantum transformation such as beam splitters and squeezers, as well as classical feedforward operations \cite{Weedbrook2012}.

In a work by Takeoka and Guha \cite{Takeoka2014} it was proven that the maximal mutual information for coherent-state encoding under the ``Gaussian receiver assumption'' is achieved using solely homodyne or heterodyne detection.
This implies that any combination of beam splitter, squeezing and classical feedforward operations will not increase the mutual information.
This conclusion holds for all possible coherent-state code words.
A generalisation of the above work including photon-counting, etc., that is non-Gaussian operations, is provided by Rosati et al.\ \cite{Rosati2016, Rosati2017}.

As an interesting curiosity, in this paper we investigate the mutual information attained by a Gaussian receiver using a simple code word consisting of phase conjugated coherent states \cite{Cerf2001} and joint detection in an amplified channel.
It was found by Niset et al.\ that a pair of phase conjugated coherent states contains more information than a pair of identical states when the quality of the information estimation is quantified by the mean fidelity \cite{Niset2007}.
It was realised that the superiority of phase conjugation is revealed through Gaussian entangled measurements corresponding to a single beam splitter followed by two homodyne detectors, and the effect has been used for improved quantum cloning \cite{Sabuncu2007}.
Similar results have been discussed for time-reversed qubits \cite{Gisin1999}.
Although a superiority of phase conjugation and joint measurements was found in these works, the conclusions only hold for the estimation based on average fidelity and assuming a symmetric Gaussian distribution.

Using mutual information as the estimator and allowing for an arbitrary Gaussian distribution, joint measurements at the receiver will not improve the scheme \cite{Takeoka2014}.
However, in this paper, we show that a channel consisting of phase-insensitive amplifiers or if one restricts the alphabet to a symmetric Gaussian, then phase conjugation combined with joint measurements is superior to identical coherent states with local measurements.
We extend the analysis by considering phase-conjugated encoding in Gaussian entangled channels, and find again that this encoding is superior to squeezed light encoding assuming channels with phase-insensitive amplification.
It is, however, also important to stress that if one allows for phase-sensitive amplification and asymmetric alphabets, there is no gain in exploiting phase conjugation and joint measurements.

\section{Analysis}

\begin{figure}
  \includegraphics{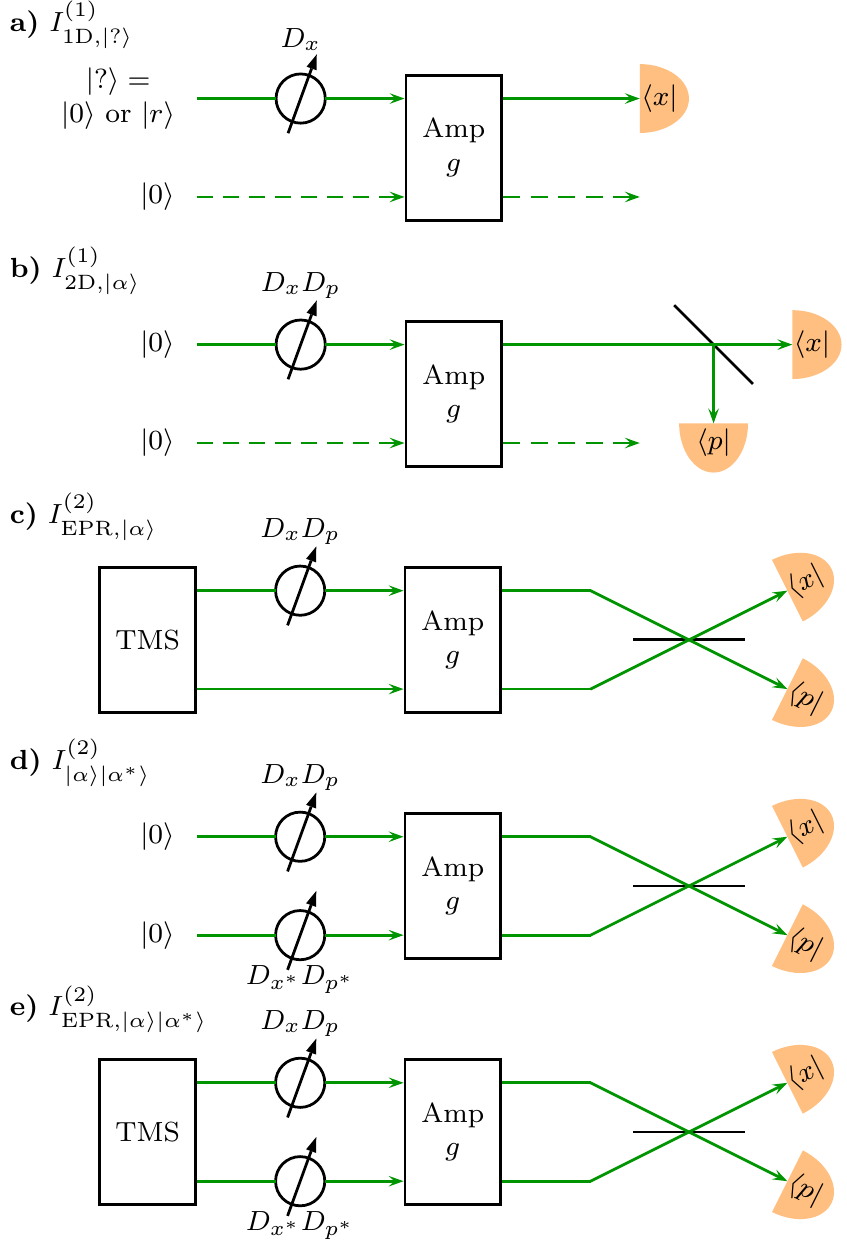}
  \caption{%
  Gaussian channels with one- and two-dimensional signal modulation $D$, amplification, and detection.
  (a) 1D alphabet, coherent state or squeezed state ($\ket r$) input.
  (b) 2D alphabet, heterodyne detection.
  (c) 2D alphabet, joint detection, two-mode squeezing (TMS).
  (d) 2D alphabet, joint detection, classical phase conjugation.
  (e) Combination of (c) and (d).
  }%
  \label{fig:scheme_all}%
\end{figure}

Let us consider a message being encoded into the displacement of a pure Gaussian state with variances $V_{\gamma, x}$ and $V_{\gamma, p}$ taken from a Gaussian distribution of variances $V_{s, x}$ and $V_{s, p}$, where the indices $x$ and $p$ are the amplitude and phase quadratures, respectively, obeying the commutation relation $[x, p] = 2 i$.
For coherent states $V_{\gamma, x} = V_{\gamma, p} = 1$, while for pure squeezed states $V_{\gamma, x} < 1$ and $V_{\gamma, p} = V_{\gamma, x}^{-1}$.
The total variances are thus $V_x = V_{s, x} + V_{\gamma, x}$ and $V_p = V_{s, p} + V_{\gamma, p}$, and the average photon number reads $\photonum = (V_x + V_p) / 4 - 1 / 2$.
All channels studied in the following are depicted in \fig\ \ref{fig:scheme_all}.

We start by evaluating the mutual information of a single use of an amplifying bosonic channel based on three different types of Gaussian alphabets followed by either homodyne or heterodyne detection:
(i) A one-dimensional (1D) coherent-state alphabet $\{P(x), \ket{\alpha = x + i p}\}$ and homodyne detection, where $P(x)$ is a Gaussian distribution with variances $V_x = V_s + 1$ and $V_y = 1$.
(ii) A 1D squeezed state alphabet $\{P(x), S(r) \ket \alpha \}$ and homodyne detection, where $S(r)$ is the squeezing operator, $r = -\ln V_{\gamma, x}$ is the squeezing parameter, and the alphabet has the variances $V_x = V_s + V_{\gamma, x}$ and $V_p = 1 / V_{\gamma, x}$.
(iii) A two-dimensional (2D) coherent-state alphabet, $\{P(\alpha), \ket \alpha\}$, with variances $V_x = V_p = V_s + 1$ and heterodyne detection.
For these three cases, depicted in \fig\ \ref{fig:scheme_all}(a)-(b), one easily finds the mutual information [which for Gaussian states is defined in terms of the noise $N$, and the signal $S$: $I = \nicefrac 1 2 \log(1 + \nicefrac S N)$] \cite{Holevo1999, Caves1994, Yamamoto1986},
\begin{align}
  I_{\text{1D}, \ket \alpha}^{(1)} &= \log(1 + 4 \photonum) / 2 \\
  I_{\text{1D}, \ket r}^{(1)} &= \log(1 + 2 \photonum) \\
  I_{\text{2D}, \ket \alpha}^{(1)} &= \log(1 + \photonum) \label{eq:I1}
\end{align}
under the photon number constraints.
Note that photons are produced both as a result of the displacement and the squeezing operation and, in case of a squeezed state alphabet, the optimal squeezed state variance is $V_{\gamma, x} = 1 / (2 \photonum + 1)$.
Next, we consider the change in the mutual information under phase-insensitive amplification, which can be described by the Bogoliubov transformations $a_\text{out1} = \sqrt g a_\text{in1} + \sqrt{g - 1} a_\text{in2}^\dagger$ and $a_\text{out2} = \sqrt g a_\text{in2} + \sqrt{g - 1} a_\text{in1}^\dagger$ where $a_\text{in/out}$ represent the input/output annihilation operators of the modes 1 and 2, and $g$ is the gain parameter \cite{Caves1982}.
Under such amplification, the mutual information of the three channels in \eqn\ \eqref{eq:I1} will change to
\begin{align}
  I_{\text{1D}, \ket \alpha}^{(1)}(g) &= \log\robra*{1 + \tfrac{4 g \photonum}{2 g - 1}} / 2,
  \lim_{g \to \infty} = \log(1 + 2 \photonum) / 2 \\
  I_{\text{1D}, \ket r}^{(1)}(g) &=
  \log\robra*{1 + \frac{4 g (\photonum + \nicefrac 1 2) - g V_\gamma - g / V_\gamma}{g V_\gamma + g - 1}} / 2 \label{eq:I2b} \\
  I_{\text{2D}, \ket \alpha}^{(1)}(\cancel g) &= I_{\text{2D}, \ket \alpha}^{(1)} \label{eq:I2}
\end{align}
where the optimal squeezing variance in \eqn\ \eqref{eq:I2b} is $V_\gamma = (g + \sqrt{g^2 - (1 - g - 4 g (\photonum + \nicefrac 1 2))(g - 1)})/(g + 4 g (\photonum + \nicefrac 1 2) - 1)$.
It is interesting to note that a linear phase-insensitive amplifier has no effect on the mutual information associated with a symmetric distribution of coherent states, while it has a degrading effect on the 1D distributions.
To highlight the amplification invariance of a channel with an amplifier, the notation $\cancel g$ is used.
The intuitive explanation is that the noise added by amplification is shared by both signal quadratures and thus its effect on the individual detection is decreased.
However, the 1D distributions can also be amplified without any effects on the mutual information by using phase-\emph{sensitive} amplifiers in place of the phase-insensitive amplifier.

\begin{figure}
  \includegraphics{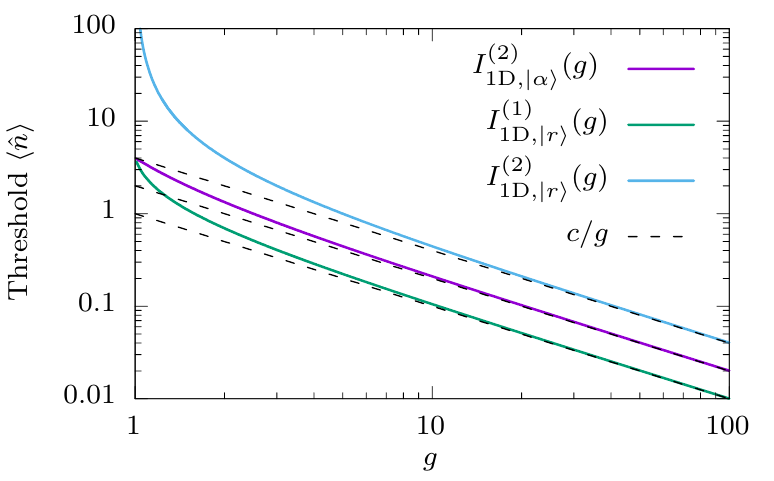}
  \caption{%
  Comparison of $I_{\text{2D}, \ket \alpha}^{(2)}$ to three amplification-variant channel-uses: \eqn\ \eqref{eq:I2b}, \eqref{eq:I4a}, and \eqref{eq:I4b}.
  The vertical axis shows the threshold \photonum\ that needs to be overcome to make $I_{\text{2D}, \ket \alpha}^{(2)}$ beneficial over the other channel use.
  Dashed lines follow the asymptotic behaviour.
  The constant factors, from top to bottom, read $c = 4, 2, 1$.
  Compared to the other two channels, $I_{\text{1D}, \ket r}^{(2)}(g)$ always outperforms $I_{\text{2D}, \ket \alpha}^{(2)}$ for $g = 1$ and thus has no finite threshold \photonum\ at this value.
  }%
  \label{fig:app_1}%
\end{figure}

We now consider two usages of the channel applying the same alphabets as above and assuming that the same state is sent twice (or two channels are employed).
That is, the three alphabets now read $\{P(x), \ket{\alpha}^{\otimes 2}\}$, $\{P(x), (S(r) \ket{\alpha})^{\otimes 2}\}$ and $\{P(\alpha), \ket{\alpha}^{\otimes 2}\}$, which results in the mutual information 
\begin{align}
  I_{\text{1D}, \ket \alpha}^{(2)} &= \log(1 + 2 \photonum) \label{eq:I3} \\
  I_{\text{1D}, \ket r}^{(2)} &= 2 \log(1 + \photonum) \\
  I_{\text{2D}, \ket \alpha}^{(2)} &= 2 \log(1 + \photonum / 2)
\end{align}
assuming homodyne detection for the 1D distributions and heterodyne detection for the 2D distributions.
For an amplifying phase-insensitive channel, we find
\begin{align}
  I_{\text{1D}, \ket \alpha}^{(2)}(g) &= \log\robra*{1 + \tfrac{2 g \photonum}{2 g - 1}},
  \lim_{g \to \infty} = \log(1 + \photonum) \label{eq:I4a} \\
  I_{\text{1D}, \ket r}^{(2)}(g) &=
  \log\robra*{1 + \frac{2 g (\photonum + 1) - g V_\gamma - g / V_\gamma}{g V_\gamma + g - 1}} \label{eq:I4b} \\
  I_{\text{2D}, \ket \alpha}^{(2)}(\cancel g) &= I_{\text{2D}, \ket \alpha}^{(2)} \label{eq:I4}
\end{align}
where the optimal squeezing variance in \eqn\ \eqref{eq:I4b} is $V_\gamma = [g + \sqrt{g^2 - [1 - g - 2 g (\photonum + 1)](g - 1)}] / [g + 2 g (\photonum + 1) - 1]$.
Again we note that the mutual information of the 1D distributions remain invariant with respect to phase-sensitive amplification.
The benefits of the amplification invariance of $I_{\text{2D}, \ket \alpha}^{(1)}$ are shown in \fig\ \ref{fig:app_1}.

Finally we consider the effect of using phase-conjugated quantum \emph{and} classical correlations between the two channels.
Three cases will be treated.
We consider
(i) phase-conjugated quantum correlations via entanglement,
(ii) phase-conjugated classical correlations using phase-space displacements, and
(iii) combined classical and quantum phase conjugation.
For all these schemes, the phase-space distributions of states (either coherent states or entangled states) are two-dimensional symmetric Gaussian.

In the first case, we assume that the correlations between the channels are of quantum origin and that they are generated by a symmetric two-mode squeezed state for which $a_\text{out1} = \cosh(r) a_\text{v1} + \sinh(r) a_\text{v2}^\dagger$ and $a_\text{out2} = \cosh(r) a_\text{v2} + \sinh(r) a_\text{v1}^\dagger$, where $a_\text{v}$ is the annihilation operator associated with vacuum and the indices refer to channels 1 and 2.
Information is encoded as symmetric Gaussian displacements onto one of the modes [\fig\ \ref{fig:scheme_all}(c)] and the states are measured using a continuous-variable Bell measurement consisting of a symmetric beam splitter and two homodyne detectors.
The channels are jointly amplified by using the two inputs of a phase insensitive amplifier.
For such quantum correlated channels, we find the mutual information
\begin{equation}
  I_{\text{EPR}, \ket \alpha}^{(2)}(\cancel g) = \log\robra*{1 + \photonum + \photonum^2 / 2}, \label{eq:I5}
\end{equation}
where the optimal two-mode squeezed state has a variance of $V = 1 / (1 + \photonum)$.
Note that this scheme is similar to continuous-variable dense coding, except for the fact that dense coding disregards the photons in the nondisplaced mode (as this mode can be sent off-line) \cite{Braunstein2000, Ralph2002}.
For dense coding, $I_\text{dc} = \log(1 + \photonum + \photonum^2)$, which is optimised for a two-mode squeezed variance of $V = 1 / (2 \photonum + 1)$.

\begin{figure}
  \includegraphics{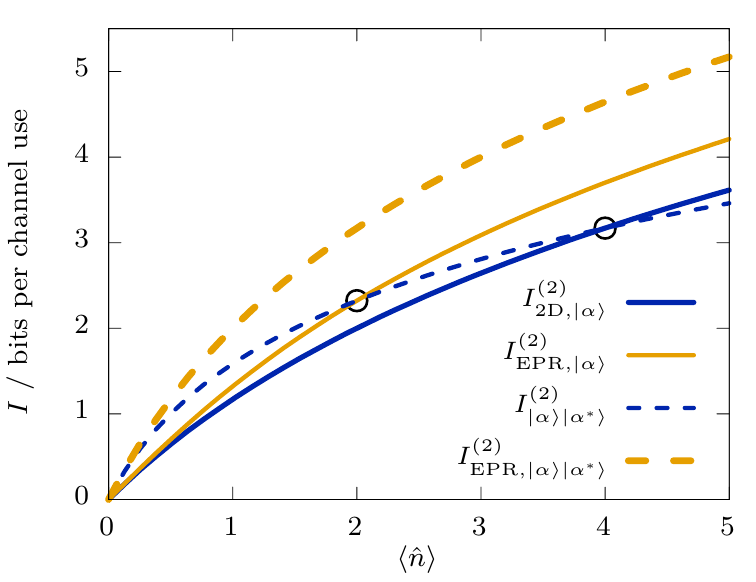}
  \caption{%
  Mutual information for 2D symmetric alphabets with heterodyne or joint detection.
  All combinations are invariant under amplification.
  Channel $I_{\text{EPR}, \ket \alpha \ket{\alpha^*}}^{(2)}$ performs best.
  Up to $\photonum = 2$, $I_{\ket \alpha \ket{\alpha^*}}^{(2)}$ beats the EPR state.
  Channel $I_{\text{2D}, \ket \alpha}^{(2)}$ outperforms $I_{\ket \alpha \ket{\alpha^*}}^{(2)}$ for $\photonum > 4$.
  }%
  \label{fig:compare_mutualinformation_1}%
\end{figure}

Now instead of using quantum correlations, we consider phase-conjugated classical correlations [\fig\ \ref{fig:scheme_all}(d)], i.e., a scheme where pairs of phase-conjugated coherent states are prepared in a Gaussian distribution, $\{P(\alpha), \ket \alpha \ket{\alpha^*}\}$.
Using the same amplifying channel as above and a joint detection strategy, we find
\begin{equation}
  I_{\ket \alpha \ket{\alpha^*}}^{(2)}(\cancel g) = \log \robra*{1 + 2 \photonum}, \label{eq:I6}
\end{equation}
and by comparing this with $I_{\text{EPR}, \ket \alpha}^{(2)}$ (see \fig\ \ref{fig:compare_mutualinformation_1}), we see that for $\photonum < 2$, the channel with classical correlations is superior to the one with quantum correlations.
Both of these schemes can, however, be beaten by a scheme that combines quantum and classical phase-conjugated correlations.
Here, phase-conjugated displacements are performed onto a two-mode squeezed state, thereby combining both types of correlations [\fig\ \ref{fig:scheme_all}(e)].
The two channels are again jointly amplified in a phase-insensitive amplifier and jointly measured using a Bell measurement.
The resulting mutual information is
\begin{equation}
  I_{\text{EPR}, \ket \alpha \ket{\alpha^*}}^{(2)}(\cancel g) = 2 \log(1 + \photonum) \label{eq:I7}
\end{equation}
and is plotted in \fig\ \ref{fig:compare_mutualinformation_1}.
It is clear that the combination of quantum and classical phase-conjugating correlations is superior to using only classical or quantum correlations.
We also notice from \eqn\ \eqref{eq:I5}, \eqref{eq:I6} and \eqref{eq:I7} that the mutual information is invariant with respect to phase-insensitive amplification, which is a result of the phase-conjugating behaviour of the amplifier and the joint measurement strategy.

\section{Discussion}

If one restricts the alphabet to a two-dimensional symmetric Gaussian distribution, it is clear that the schemes consisting of phase conjugation and joint measurements ($I_{\ket \alpha \ket{\alpha^*}}^{(2)}$ and $I_{\text{EPR}, \ket \alpha \ket{\alpha^*}}^{(2)}$) are superior to a scheme based on identical states ($I_{\text{2D}, \ket \alpha}^{(2)}$).
This is also seen from \fig\ \ref{fig:compare_mutualinformation_1} where the mutual information associated with the different symmetric Gaussian distributions is plotted.

Based on these observations, it appears that the joint detection strategy is superior to independent homodyne or heterodyne measurements.
How does this conclusion comply with the result of Takeoka and Guha \cite{Takeoka2014} where it was found that the optimum mutual information is obtained simply by individual homodyne or heterodyne detection?
There is no contradiction since our conclusion is based on the assumption that we are using symmetric Gaussian distributions and thereby not allowing for asymmetric Gaussian distributions.
It is clear from \eqn\ \eqref{eq:I3} that by using coherent states encoding a 1D Gaussian alphabet and individual homodyne detection, the mutual information is identical to the channel with phase-conjugated coherent states [\eqn\ \eqref{eq:I6}].
Likewise, we see that the mutual information of the entangled and phase-conjugated channel [\eqn\ \eqref{eq:I7}] can be also reached with individually squeezed states and direct homodyne detection (\eqn\ \eqref{eq:I3}).
Therefore, the mutual information associated with independent Gaussian measurements cannot be beaten, in agreement with Takeoka and Guha \cite{Takeoka2014}.

\begin{figure}
  \includegraphics{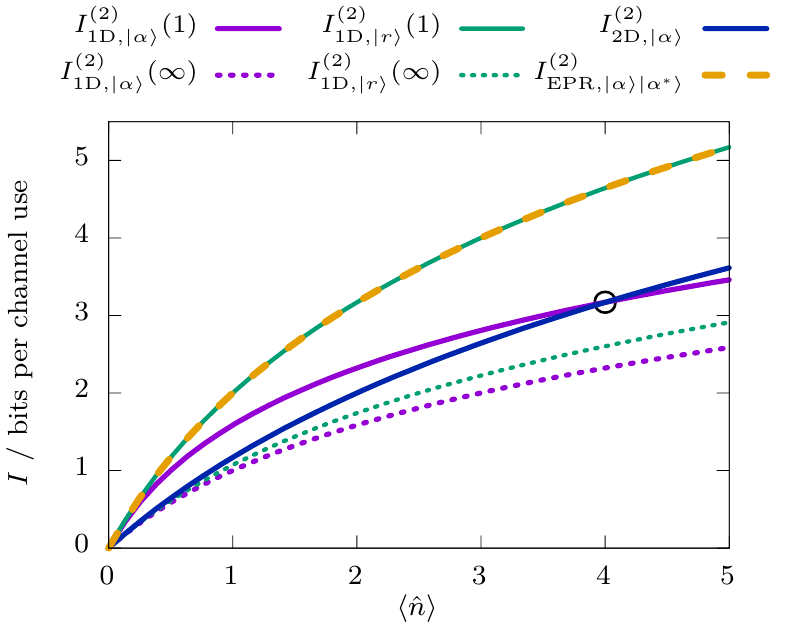}
  \caption{%
  Mutual information of a phase-insensitive amplifying channel for four different alphabets.
  Out of these four, two alphabets are variant under gain and the mutual information for $g = 1$ and $g \to \infty$ is shown.
  The largest mutual information is reached for the quantum and classical phase-conjugating alphabet, $I_{\text{EPR}, \ket \alpha \ket{\alpha^*}}^{(2)}$, and for $I_{\text{1D}, \ket r}^{(2)}$.
  However, $I_{\text{1D}, \ket r}^{(2)}$ suffers significantly from amplification, in contrast to $I_{\text{EPR}, \ket \alpha \ket{\alpha^*}}^{(2)}$ which is invariant under amplification.
  Double use of a 1D alphabet ($I_{1D, \ket \alpha}^{(2)}$) outperforms a double use of a 2D alphabet until $\photonum = 4$.
  Under amplification, the advantage over $I_{\text{2D}, \ket \alpha}^{(2)}$ vanishes, which highlights the benefits of schemes invariant under amplification.
  }%
  \label{fig:compare_mutualinformation_2}%
\end{figure}

Another restriction that leads to a superiority of phase-conjugation and joint measurements is when linear phase-insensitive amplification is performed.
Here, the intrinsic phase-conjugation process of the amplifier leads to state amplification without any information degradation resulting from the joint measurement strategy.
This holds for both pure classical phase conjugation [\eqn\ \eqref{eq:I6}], pure quantum phasec-conjugated correlations (\eqn\ \eqref{eq:I5}), and for combined phase-conjugated classical and quantum correlations (\eqn\ \eqref{eq:I7}).
In contrast, using phase-insensitive amplification in the channels with identical coherent states or squeezed states, the information will be degraded as deduced in \eqn\ \eqref{eq:I4a} and \eqref{eq:I4b}.
The comparison is shown in \fig\ \ref{fig:compare_mutualinformation_2}.
However, as noted above, by using phase-sensitive amplifiers instead of a phase-insensitive amplifier, the mutual information of the individual squeezed or coherent states (with individual detection) is unchanged.

\section{Conclusion}

In conclusion, we have found that a phase-conjugated correlated alphabet combined with a continuous-variable Bell measurement in a Gaussian channel is superior to using identical states and individual Gaussian detection if the alphabet is restricted to a symmetric Gaussian distribution or if the channel is amplified symmetrically in phase space corresponding to a phase-insensitively amplifying channel.
Moreover, we identified cases where amplification has no effect on the mutual information in contrast to the standard scheme where amplification degrades the information.
The study has been limited to single- and double-channel encoding, but as an outlook it would be intriguing to consider phase-conjugated and multipartite entangled encoding in an arbitrary number of channels thereby generalising the analysis.

\paragraph{Acknowledgement}

We acknowledge support from the Danish National Research Foundation, grant DNRF142, and the Innovation Fund Denmark.
Also, we thank Masahiro Takeoka for interesting discussions.

\setcounter{equation}{0}
\def\theequation{A\arabic{equation}}

\section{Appendix}

The high gain limit for \eqn\ \eqref{eq:I2b} is
\begin{equation}
  I_{\text{1D}, \ket r}^{(1)}(\infty) = \log \left(\tfrac{\sqrt{n + 1} (4 n + 3)^2}{5 \sqrt{n + 1} + 4 n \left(\sqrt{n + 1} + 1\right) + 4}\right) / 2
\end{equation}
and for \eqn\ \eqref{eq:I4b}
\begin{equation}
  I_{\text{1D}, \ket r}^{(2)}(\infty) = \log \left(\tfrac{\sqrt 2 \sqrt{n + 2} (2 n + 3)^2}{2 \sqrt{2} \sqrt{n + 2} n + 4 n + 5 \sqrt{2} \sqrt{n + 2} + 8}\right).
\end{equation}

\end{document}